# Mode coupling theory of electrolyte dynamics: Time dependent diffusion, dynamic structure factor and solvation dynamics

**Susmita Roy, Subramanian Yashonath, Biman Bagchi***

Solid State and Structural Chemistry Unit, Indian Institute of Science, Bangalore 560012, India.

*Email: bbagchi@sscu.iisc.ernet.in

***ABSTRACT***

**A self-consistent mode coupling theory (MCT) with microscopic inputs of equilibrium pair correlation functions is developed to analyze electrolyte dynamics. We apply the theory to calculate concentration dependence of (i) time dependent ion diffusion, (ii) dynamic structure factor of the constituent ions, and (iii) ion solvation dynamics in electrolyte solution. Brownian dynamics (BD) with implicit water molecules and molecular dynamics (MD) method with explicit water are used to check the theoretical predictions. The time dependence of ionic self-diffusion coefficient and the corresponding dynamic structure factor evaluated from our MCT approach show quantitative agreement with early experimental and present Brownian dynamic simulation results. With increasing concentration, the dispersion of electrolyte friction is found to occur at increasingly higher frequency, due to the faster relaxation of the ion atmosphere. The wave number dependence of total dynamic structure factor $F(k, t)$, exhibits markedly different relaxation dynamics at different length scales. At small wave numbers, we find the emergence of a step-like relaxation, indicating the presence of both fast and slow time scales in the system. Such behaviour allows an intriguing analogy with temperature dependent relaxation dynamics of supercooled liquids. We find that solvation dynamics of a tagged ion exhibits a power law decay at long times – the decay can also be fitted to a stretched exponential form. The emergence of the power law in solvation dynamics has been tested by carrying out long Brownian dynamics simulations with varying ionic concentrations. This solvation time correlation and ion-ion dynamic structure factor indeed exhibits highly interesting, non-trivial dynamical behaviour at intermediate to longer times that require further experimental and theoretical studies.**

# I. Introduction

Theory of ionic conductivity in electrolyte solutions is a historically important problem that forms the very core of chemistry of solutions [1-11]. It is important to understand the behaviour of different electrolytes as they are omnipresent in nature and serves as an essential ingredient of chemical industry. The most abundant sources of electrolytes are of course ocean and sea water. Ocean water is an electrolyte solution comprising about 0.5 mol L$^{-1}$ of sodium chloride and significant concentrations of other ions and has pH typically limited to a range between 7.5 and 8.4 [12]. Probably due to the fact that life evolved in nature in the presence of electrolytes, a biological cell contains a large number of electrolytes with ions of various kinds.

A complex electrolyte balance is always crucial for the maintenance of many biological and chemical activities in water [12-16]. Precise concentration gradients of different electrolytes between intracellular and extracellular milieu of higher organisms affect and regulate the hydration of the body, blood pH, and are also critical for nerve and muscle functions. The tightly controlled concentration dependence of electrolyte dynamics is extremely important for life processes [17]. The dynamical impact they exert and the complex interaction with other common molecules like water, also help to understand their role in different physical, chemical and biological processes. Thus the dynamics of electrolyte solutions remains and shall remain a central area of research in physical chemistry for a long time to come.

Despite many attempts to understand the complex behaviour of electrolyte solutions over many decades, a number of fundamental problems still remain unsolved. Among several other complexities, the long range nature of ion-ion and ion-solvent interactions have hindered theoretical and computational advancement [2]. To deal with dynamics of such complex liquids, the relevant motion, whether molecular, or of a collection of atoms, is often described in terms of a generalized Langevin equation [18-21]. For many such applications, we need to know frequency dependence of friction experienced by a moving ion. Due to the presence of both short range order and long range correlations, electrolyte dynamics is found to be bimodal/biphasic in nature. The rapid short time response in aqueous electrolyte solutions arises partly due the interaction with surrounding water molecules while the slow long time response arises due to the collective dynamics involving many particles. The latter is descibed in terms of ion atmosphere relaxation.

Beyond such wide separation of time scales, what more amazing is that many liquids show an initial response whose rate is many orders of magnitude greater than that of long time response of the same liquid. Sometimes it is found that decay in the intermediate time range can become even slower than the long time decay. To understand the origin of such wide range of timescales (beginning from a few tens of femtoseconds for the initail response to minutes even extending to hours for the stress relaxation in glassy liquids) a general theoretical framework known as Mode coupling theory (MCT) has been developed.

MCT can be best viewed as a synthesis of two formidable theoretical approaches, namely the renormalized kinetic theory of liquids and the extended hydrodynamic theory of fluids [22-25]. However none of these two approaches provides a self-consistent description to understand coexistence of these short and long time response of liquid state dynamics [25]. The self-consistency enters in the determination of time correlation functions of the hydrodynamic modes in terms of transport coefficient/memory function, and then use of these time correlation functions to obtain the transport coefficients. It is worth mentioning here that the first self-consistent calculation was done by Geszti [26] to describe the growth of viscosity in a previtrification region. Later Leutheusser [27], Kirkpatrick [28], Gotze et al. [29, 30] performed self-consistent calculations to explain the divergence of dynamic structure factor and viscosity near glass transition. As our study is focused towards the investigation of diffusion and some other dynamical quantities associated with the chemical dynamics of electrolyte solutions, only certain aspects along with the success and failure of MCT approch will be discussed.

MCT approach provides a simple and microscopic derivation of the elementary equations or, laws of electrolyte transport. The first important development is the Debye-Huckel-Onsager law of concentration dependence of ionic conductivity [31,32].

$$\Lambda_\alpha(c_\alpha) = \Lambda_\alpha^0 - [A + B\Lambda_\alpha^0]\sqrt{c_\alpha} \qquad (1)$$

where $\Lambda_\alpha$ is the conductance of the ionic species $\alpha$ at concentration $c_\alpha$. $\Lambda_\alpha^0$ is the conductance at infinite dilution [33]. A and B are the two constants that are determined by the properties of ions and the medium. For a binary electrolyte, the complete expression of these constants are given in ref. 33.

The second important advancement is the derivation of the Debye-Falkenhagen-Onsager (DFO) law describing the frequency dependence of electrolyte conductance [34-36]. The early derivation of the DFO limiting law was based on irreversible thermodynamics and macroscopic hydrodynamics. The third important advancement is the derivation of Onsager-Fuoss expression of the concentration dependence of viscosity [37]. Expressions of Debye-Falkenhagen-Onsager-Fuoss (DFOF) limiting laws [38], based on the Debye-Hückel (DH) ion pair distribution functions can efficiently describe the motion of ions in an electrolyte solution in terms of specific conductivity, diffusion and other rheological transport coefficients with their distinctive square-root in concentration dependence at very low ion concentration (up to $10^{-2}$M) [39]. In addition, these laws provide elegant explanations of the different experimental observations on ion diffusion constant reported by QENS and NMR measurements [40].

In experimental situations, one often encounters aqueous electrolyte solutions with varying concentration. In concentrated solutions when ion-ion correlations play an important role, a quantitative description of the role of electrolyte solution on various physico-chemical processes appears to be a formidable problem. For instance, dynamics of DNA or protein is greatly influenced by the surrounding ion and water molecules. In recent times, several time dependent fluorescence Stokes shift (TDFSS) measurements revealed a surprising slow anomalous power decay in the solvation dynamics of DNA [41-45]. This power-law kinetics appears similar to the dynamics of proteins than the solvation dynamics of simple liquids. To study DNA solvation dynamics in TDFSS experiments, a suitably placed dye molecule is excited optically and the shift in the frequency of the emission spectrum is measured as a function of time. In such experiment either a base of the DNA duplex is replaced biochemically by a dye molecule, such as coumarin or 2-aminopurine [46], or, a fluorescent probe is intercalated to one of the grooves of the DNA [41-45]. However in both these approaches the existence of an apparently power law decay component is common. Note here that such an interesting behaviour in DNA solvation was first reported by Berg and co-workers where they showed a logarithmic time dependence (ranged from 40 ps to 40 ns) of the solvation dynamics using coumarin as a probe [41-44]. The short time contribution is dominated by ion-water coupling and the decay time constant is less than about 10 ps [41-44, 47]. However, Zewail and co-workers measured the time-resolved Stokes shift of 2-aminopurine in an oligonucleotide over 100 fs to 50 ps time range and the relaxation behavior was found to be biexponential [46]. While the faster component was assigned to unperturbed water and the slower one was attributed to "biological water". One can then imagine that even when structured water close to DNA groove could be involved, the slow decay might not prolong beyond 100 ps and may not give rise to a power law with such a small power as reported by Berg and co-workers [41-44]. This argument suggests that motion of components other than water, such as ions, DNA backbone, and DNA bases may have an important role in the solvation response. A pertinent explanation of the issue of the power law is still not fully elucidated.

In a recent study Bagchi suggested that a slow decay may originate from collective ion contribution of the buffer solution because in the solvation dynamics experiments of DNA, DNA is invariably immersed in a buffer solution [40]. The complex interplay between long range probe-ion direct correlation function and ion-ion dynamic structure factor are also liable for such anomalous behavior. Such power law decay, however, has not been reflected in any early computer simulation studies where run time was limited within the time range of 1ns or so[47]. It seems to appear in the longer timescale.

In general solvation dynamics is explained by the decay of the non-equilibrium TCF which is denoted by $S(t)$ and expressed as follows,

$$S(t) = \frac{\bar{v}(t) - \bar{v}(\infty)}{\bar{v}(0) - \bar{v}(\infty)} \qquad (2)$$

where $\bar{v}(0)$, $\bar{v}(t)$, $\bar{v}(\infty)$ represent average observed emission frequencies at the respective time zero, $t$ and infinity ($\infty$). The averages are taken over the entire emission spectrum.

An alternative, rather simpler to implement in computer simulation studies, is to study the linear response solvation dynamics from equilibrium time correlation functions. This nonequilibrium TCF can be related to equilibrium TCF of energy fluctuation by linear response (LR) theory. The equilibrium TCF is denoted by $C_S(t)$ and expressed as follows:

$$C_S(t) = \frac{\langle \delta E(0) \delta E(t) \rangle}{\langle \delta E(0) \delta E(0) \rangle} \qquad (3)$$

where $\delta E(t) = E(t) - \langle E \rangle$ represents the fluctuation of potential energy of the probe molecule with the rest of the simulation system at time t and $\langle ... \rangle$ corresponds to the equilibrium ensemble average of its ground/excited state. While $S(t)$ holds the characteristics of the initial state to that of the final excited state of the probe, $C_S(t)$ characterizes the ground state. However in most of the cases this difference along with the difference of $C_S(t)$ from its excited state are ignored.

Now to understand the role of ions in the solvation dynamics experiments we start with the following expression of solvation energy of a probe located at position **r** and at time t, provided by the time dependent density functional theory of statistical mechanics,

$$E_{sol}(\mathbf{r},t) = -k_B T n_{ion}(r,t) \int d\mathbf{r}' \, d\Omega' c_{p,i}(\mathbf{r},\mathbf{r}',\Omega') \delta \rho_i(\mathbf{r}',\Omega',t), \qquad (4)$$

where $c_{p,i}$ is the probe - ion direct correlation function and $n_{ion}(r,t)$ is the position dependent number density of the ion i. $\delta \rho_i(\mathbf{r}',\Omega',t)$ is the fluctuation in the position (**r**), orientation ($\Omega$), and time (t) dependent number density of the ionic solution. This above equation leads to a microscopic expression of solvation energy that includes the effects of self-motion of the ion.

Consequently, the solvation energy-energy time correlation function (TCF) which forms the notion of our study of solvent dynamics of a mobile solute ion can be given as follows,

$$<\delta E(0)\delta E(t)> = \int dk\, k^2 \sum_{i,j=1}^{2} c_{pi}(k) c_{pj}(k) F_{ij}(k,t) + \int dk\, k^2 c_{p-sol}^2(110,k) F_{sol}(110,k,t)$$
$$+ \int dk\, k^2 c_{p-sol}^2(111,k) F_{sol}(111,k,t) \quad (5)$$

where $c_{pi}(k)$ is the Fourier transform of i-th ion-probe wave number dependent direct correlation function, $c_{p-sol}^2(11m,k)$ are the probe-solvent direct correlation function, expanded into spherical harmonics, $F_{ij}(k,t)$ are the partial intermediate scattering function between ion of type i and of type j. $F_{sol}(11m,k,t)$ are the spherical harmonic expansion of the angle dependent intermediate scattering function of surrounding solvent [48].

This complex expression can be simplified in the long time limit because the last two terms in the above expression decays on a much faster time scale as they involve rotational motion of water molecules. It is important to mention here that water molecules even in the grooves of DNA rotate rather fast, on the time scale of tens of ps. These motions are of course important to understand solvation dynamics in the ultrafast (sub-ps) to intermediate time scales (of the order of tens of ps) but not expected to make any significant contribution to the slow times (of the order of ns) where the power decay appears. We therefore simplify the above expression by keeping only the first term that involves slow motion of ions.

Now, we assume that the probe can be approximated by a point dipole. Under this approximation, $c_{p+}(k)$ and $c_{p-}(k)$ are the same. We can then merge the two terms

$$<\delta E(0)\delta E(t)> = A[\int_0^\infty dk\, k^2 c_{p+}^2(k) F_{++}(k,t) + \int_0^\infty dk\, k^2 c_{p-}^2(k) F_{--}(k,t) + 2\int_0^\infty dk\, k^2 c_{p+}(k) c_{p-}(k) F_{+-}(k,t)]$$

(6)

where A is a numerical constant. While the ion-dipole direct correlation function is long ranged, it is the evaluation of the intermediate scattering function $F_{++}(k,t)$ of the ions that requires special attention for the characterization of solvation time correlation function..

While computer simulation investigation of DNA solvation dynamics at a molecular level and at longer timescale is still a challenge, ion atmosphere relaxation in the electrolyte solution that is coupled to solvation dynamics can be probed theoretically by the classy approach as mentioned before, i.e. Mode coupling theory [49-51].

The above experimental findings greatly encouraged us first to investigate an electrolyte solution and the corresponding dynamical behavior at different time scales. It is worth mentioning here that while Debye-Huckel-Onsager-Falkenhagen theory of the low concentration limiting laws explaining ionic transport properties has been regarded as one of great achievements of nonequilibrium statistical mechanics there still exists a lacuna in that this theory does not explain the mobility of an individual ion at zero concentration in terms of the ion-solvent interaction. In this work we are more concerned about understanding the single ion mobility in terms of self-diffusion constant. There are now compelling reasons for developing a microscopic theory of self-diffusion of concentrated electrolyte solutions. We exploit the well-known mode coupling theory (MCT) that has been extended recently to treat electrical conductivity [10,11].

## II. Theoretical formulation

### A. MCT Approach and choice of slow variables

The basic concept of MCT is that the fluctuation of a given dynamical variable decays predominantly into pairs of hydrodynamic modes associated with conserved single-particle or collective dynamical variables. Effectively in hydrodynamic approach the liquid-state dynamics is based on the assumption that many experimental observables (like the intensity in a light scattering experiment) can be rationalized by considering the dynamics of a few slow variables [52]. The natural choice for the slow variables are the densities of the conserved quantities such as, number density, momentum density, and energy density. The conservation of number, eneregy, and momentum are locally expressed by the conservation laws [53]. In MCT, ion-ion dynamical correlation terms are usually determined by using two slow variables: charge density and charge current. These essentially determine electrolyte dynamics. While they offer small corrections to viscosity, their effect on electrolyte conductivity is very large.

### B. Basic model and definitions

Let us begin with an electrolyte solution comprising of positive and negative ions immersed in a continuum solvent of dielectric constant, $\varepsilon$. The ions interact through a spherically symmetric short-range potential and a long-range Coulombic interaction potential which is scaled by the value of the dielectric constant. The pair potential of interaction between two ions of charge $q_\alpha$ and $q_\beta$ is expressed as

$$U_{\alpha,\beta}(r) = U_{\alpha\beta}^{SR}(r) + \frac{q_\alpha q_\beta}{\varepsilon r} \tag{7}$$

where $r$ is the distance between the two ions and $U_{\alpha\beta}^{SR}(r)$ stands for a spherically symmetric short-range interaction potential. We express the position (r) and time ($t$) dependent number density of species $\alpha$ as $\rho(r,t)$ and its Fourier transform $\rho(k,t)$ is defined by,

$$\rho_\alpha(k,t) = \int_{-\infty}^{\infty} d\mathbf{r}\, e^{i\mathbf{k}\cdot\mathbf{r}} \rho_\alpha(\mathbf{r},t) \tag{8}$$

The Fourier transform of the ion-ion intermediate scattering function, $G_{\alpha\beta}(\mathbf{k},t)$ is given by,

$$G_{\alpha\beta}(\mathbf{k},t) = (N_\alpha N_\beta)^{-1/2} \langle \rho_\alpha(\mathbf{k},t)\rho_\beta(-\mathbf{k},t) \rangle \tag{9}$$

where <...> denotes average over an equilibrium ensemble. $N_\alpha$ and $N_\alpha$ are, respectively, the number of ions of species $\alpha$ and $\beta$ in the solution. Finally, the Fourier-Laplace transform $G_{\alpha\beta}(\mathbf{k},z)$ is defined by,

$$G_{\alpha\beta}(\mathbf{k},z) = \int_0^\infty dt\, e^{-zt} G_{\alpha\beta}(k,t) \tag{10}$$

We define $v_s(t)$ as the time-dependent velocity of a single tagged ion of charge, $q_s$. Its time evolution can be described by the following generalized Langevin equation:

$$\frac{\partial}{\partial t} v_s(t) = -\int_0^\infty dt\, \zeta_s(t-t') v_s(t') + f_s(t) \tag{11}$$

where $\zeta_s(t)$ is the total friction, $f_s(t)$ acting on the single tagged ion and is the so-called random force. The frequency-dependent friction is $\zeta_s(z)$ defined as the Laplace transform of $\zeta_s(t)$. The self-diffusion coefficient $D_s(z)$ is related to the friction $\zeta_s(z)$ by the following generalized Einstein relation,

$$D_s(z) = \frac{k_B T}{m}\left[z + \zeta_s(z)\right]^{-1} \quad (12)$$

where $k_B$ is Boltzmann constant, T is the temperature, and $m$ is the mass of the tagged ion. In the present work, other than frequency dependence of friction and diffusion, our focus will be on the calculation of the zero frequency friction $\zeta_s(z=0)$ and the zero frequency self-diffusion coefficient, $D_s(z=0)$.

### C. Total electrolyte friction

The total friction acting on the tagged ion now can be decomposed into two parts. The first part is caused by the microscopic interaction of the tagged ion with the surrounding solvent molecules and ions and the second part originates from the hydrodynamic coupling of the velocity of the tagged ion with the current modes of the surrounding particles. Thus, the total friction on the tagged ion can be expressed as,

$$\frac{1}{\zeta(z)} = \frac{1}{\zeta_{s,mic}(z)} + \frac{1}{\zeta_{s,hyd}(z)} \quad (13)$$

The above equation can be easily analyzed [51, 53-55]. A tagged ion diffuses by two mechanisms that are as follows: The first one is by the random walk caused by its interactions with the surrounding solvent and ion molecules. The second is the random walk caused by the natural currents or flows present in the liquid. These two contributions to diffusions are additive, as they originate from two different types of motions. The mechanisms are, however, coupled at a dynamic level which, in this theory, enters nicely through self-consistency mentioned earlier. It has been shown by the mode coupling theory that a small neutral solute's diffusion in nonpolar liquid is dominated by the microscopic term only, that is, by the terms which essentially arise from collisional contributions and density fluctuations. However, the electrphoretic term is significant only at rather high frequency as it is connected with charge current density. So, for simplicity, the hydrodynamic contribution can be neglected as we are more concerned about the long time phenomena.

MCT is now used to calculate the ionic friction. Mode coupling theory (MCT) shows that the microscopic ionic friction can be decomposed in to two parts, Stokes friction and electrolyte friction

$$\zeta_{mic-ion}(z) = \zeta_{Stokes} + \zeta_{elec}(z) \qquad (14)$$

We have invoked a time scale separation inherent in writing the above expression. The expression reflects the separation of time scales between ion atmosphere relaxation and local density (and also momentum which has even faster) relaxation.

Here we have neglected the frequency dependence of the Stokes friction due to viscosity [53]. It has been often shown by the mode coupling theory that a small neutral solute's diffusion in nonpolar liquid is dominated by the microscopic term only, that is, by the terms which arise from collisional contributions and density fluctuations. $\zeta_{elec}(z)$ is the concentration-dependent electrolyte friction which originates from microscopic interaction of the tagged ion with its fluctuating ion atmosphere and is the subject of the present investigation. We shall refer to this term as the microscopic electrolyte friction.

D. **Calculation of time-dependent microscopic friction**

The microscopic friction, on the other way, is most easily analysed by using the Kirkwood's formula for friction which expresses it in terms of integration over the force-force time correlation function [56].

$$\delta\zeta_{mic-ion}(t) = \frac{1}{3k_BT}\int d\mathbf{r}\langle F(\mathbf{r},t)F(\mathbf{r},0)\rangle \tag{15}$$

$F(r,t)$ is the time-dependent force exerted on the tagged ion due to its interaction with all other ions in the solution. We obtain by using the well-known time-dependent density functional theory and it is given by,

$$F(\mathbf{r},t) = k_BT\, n_s(\mathbf{r},t)\nabla\sum_{\alpha}\int d\mathbf{r}' C_{sa}(\mathbf{r},\mathbf{r}')\delta\rho_{\alpha}(\mathbf{r}',t) \tag{16}$$

where $n_s(\mathbf{r},t)$ is the density distribution of the tagged ion, $\delta\rho_{\alpha}(\mathbf{r}',t)$ is the collective density fluctuation of species $\alpha$, and $C_{sa}(\mathbf{r},\mathbf{r}')$ is the direct correlation function between species $\alpha$ and the tagged ion. We can write the microscopic friction as an integral over wave vector space.

$$\delta\zeta_{mic-ion}(t) = \frac{1}{3k_BT(2\pi)^3}\int d\mathbf{k}\langle F(\mathbf{k},t)F(-\mathbf{k},0)\rangle \tag{17}$$

We take the Fourier transform of Eq. 11 and substitute the resulting expression in Eq. 12 to obtain the following formal expression for the electrolyte friction

$$\delta\zeta_{s,mic}(t) = \frac{k_BT}{3(2\pi)^3}\sum_{\alpha,\beta}\int d\mathbf{k}\, k^2\left[c_{sa}(k)\right]\sqrt{\rho_{\alpha}\rho_{\beta}}\left[G_{\alpha\beta}(k,t)\right]\left[c_{s\beta}(k)\right]\left[F_S(k,t)\right] \tag{18}$$

where $G_{\alpha\beta}(k,t)$ is the ionic intermediate scattering function and $F_S(k,t)$ is the self-dynamic structure factor of the tagged ion. Inverse Laplace transform of Eq. 18 provides the frequency dependence of friction of tagged ion in the solution. Here we want to emphasize that Eq. 12 and inverse Laplace transform of Eq. 18 are the two coupled equations that are elegantly associated with a self-consistency that can efficiently give zero frequency friction and diffusion. The ion-ion intermediate scattering function, $G_{\alpha\beta}(k,t)$ is given by,

$$G_{\alpha\beta}(k,t) = (N_\alpha N_\beta)^{-1/2} \langle \rho_\alpha(\mathbf{k},t) \rho_\beta(\mathbf{k},0) \rangle \tag{19}$$

where $N_\alpha$ and $N_\beta$ are, respectively, the number of ions of species $\alpha$ and $\beta$ in the solution. $\langle ... \rangle$ denotes average over an equilibrium ensemble. We represent $G_{\alpha\beta}(k,z)$ as the frequency-dependent intermediate scattering function obtained by Laplace transformation of $G_{\alpha\beta}(k,t)$. Time-dependent density functional theory is then applied that help us to obtain the following equation for the frequency-dependent intermediate scattering function

$$G_{\alpha\beta}(k,z) = \left[z + D_\alpha(z)k^2\right]^{-1} S_{\alpha\beta}(k) + \frac{D_\alpha(z)k^2}{z + D_\alpha(z)k^2} \sum_{\gamma=1}^{2} \rho_\alpha \rho_\gamma c_{\alpha\gamma} G_{\alpha\gamma}(k,z) \tag{20}$$

where the frequency-dependent diffusion coefficient $D_\alpha(z)$ is related to friction by Eq. 12. $S_{\alpha\beta}(k)$ is the partial static structure factor between species $\alpha$ and $\beta$ while $S_{\alpha\beta}(k) = G_{\alpha\beta}(k,t=0)$. $S_{\alpha\beta}(k)$ is related to the Fourier transform of the pair correlation function $h_{\alpha\beta}(k)$ by the following relation:

$$S_{\alpha\beta}(k) = \delta_{\alpha\beta} + \sqrt{\rho_\alpha \rho_\beta} h_{\alpha\beta}(k) \tag{21}$$

Here we denote the positive ions as species "1" and the negative ions as species "2" and define a intermediate scattering function matrix $G_{\alpha\beta}(k,z)$ and a structure factor matrix $S(k)$ of the following form:

$$[G(k,z)] = \begin{bmatrix} G_{11}(k,z) & G_{12}(k,z) \\ G_{21}(k,z) & G_{22}(k,z) \end{bmatrix} \tag{22}$$

and,

$$[S(k)] = \begin{bmatrix} S_{11}(k) & S_{12}(k) \\ S_{21}(k) & S_{22}(k) \end{bmatrix} \tag{23}$$

Their coupled equations as mentioned earlier can be now be expressed in a simplified form:

$$[G(k,z)] = [C(k,z)]^{-1}[S(k)] \tag{24}$$

where $[C(k,z)]$ is expressed by the following 2 X 2 matrix

$$[C(k,z)] = \begin{bmatrix} z + D_1(z)k^2[1 - \rho_1 c_{11}(k)] & -D_1(z)k^2\sqrt{\rho_1\rho_2}\, c_{12}(k) \\ -D_2(z)k^2\sqrt{\rho_1\rho_2}\, c_{21}(k) & z + D_2(z)k^2[1 - \rho_2 c_{22}(k)] \end{bmatrix} \tag{25}$$

We next solve the inverse of the $[C(k,z)]$ matrix to obtain the following explicit expression for the frequency dependence of the ionic intermediate scatteringfunctions:

$$G_{11}(k,z) = \frac{1}{\Gamma(k,z)}\left[\{z + D_2(z)k^2(1-\rho_2 c_{22}(k))\}S_{11}(k) + D_1(z)k^2\sqrt{\rho_1\rho_2}c_{21}(k)S_{21}(k)\right] \quad (26)$$

$$G_{12}(k,z) = \frac{1}{\Gamma(k,z)}\left[\{z + D_2(z)k^2(1-\rho_2 c_{22}(k))\}S_{12}(k) + D_1(z)k^2\sqrt{\rho_1\rho_2}c_{12}(k)S_{22}(k)\right] \quad (27)$$

$$G_{21}(k,z) = \frac{1}{\Gamma(k,z)}\left[\{z + D_1(z)k^2(1-\rho_1 c_{11}(k))\}S_{21}(k) + D_2(z)k^2\sqrt{\rho_1\rho_2}c_{21}(k)S_{11}(k)\right] \quad (28)$$

$$G_{21}(k,z) = \frac{1}{\Gamma(k,z)}\left[\{z + D_1(z)k^2(1-\rho_1 c_{11}(k))\}S_{21}(k) + D_2(z)k^2\sqrt{\rho_1\rho_2}c_{21}(k)S_{11}(k)\right] \quad (29)$$

where,

$$\Gamma(k,z) = z^2 + z\Delta(k)[D_1(z)k^2 S_{22}(k) + D_2(z)k^2 S_{11}(k)] + D_1(z)D_2(z)k^4\Delta(k) \quad (30)$$

and $\Delta(k) = \left[S_{11}(k)S_{22}(k) - S_{12}(k)^2\right]^{-1}$ \quad (31)

In deriving Eqs. 26-31, we have also used the following relation between $c_{\alpha\beta}(k)$ and $S_{\alpha\beta}(k)$ for a two-component system,

$$1 - \rho_2 c_{11}(k) = \Delta(k)S_{11}(k) \quad (32)$$

$$\sqrt{\rho_1\rho_2}c_{12}(k) = \Delta(k)S_{12}(k) \quad (33)$$

We note that the time dependence of the intermediate scattering functions, in terms of dynamic structure factor can be obtained through Laplace inverse transformation of frequency dependence of intermediate scattering functions. We express the Laplace transform of the self-dynamic structure factor of the tagged ion by the following equation:

$$F_s(k,z) = \frac{1}{z + D_s(z)k^2} \tag{34}$$

We still need the solutions of the static structure factors and the direct correlation functions for the calculation of the microscopic electrolyte friction. Here we note that the direct correlation functions are related to the static structure factors by Eqs. 32 and 33. The static structure factors are related to the pair correlation functions by Eq. 21. Thus, we now have to find the solutions of the pair correlation functions for the calculation of the quantities, $S_{\alpha\beta}(k)$ and $c_{\alpha\beta}(k)$ and we need to specify the nature of the short-range interaction between ions for this purpose. We consider the ions to be charged hard spheres with their diameters becoming the parameters of the model solution. Here we use the expressions derived by Phill Attard for the ionic pair correlations which are quite accurate even at high concentration limit [57]. In this scheme, the functional forms of the ionic pair correlation functions are similar to those given by Debye-Huckel theory. However, the screening parameter which enters into the mathematical expressions is not the one of Debye-Huckel theory but a renormalized one. In real space, the expression of the pair correlation function $h_{\alpha\beta}(k)$ is given by,

$$h_{\alpha\beta}(r) = -\frac{q_\alpha q_\beta}{\varepsilon k_B T (1+\kappa\sigma)} \frac{\kappa^2}{\kappa_D^2} \frac{e^{-\kappa(r-\sigma)}}{r} \tag{35}$$

where $\sigma$ is the ionic diameter, which is assumed to be the same for both kinds of ions of the solution. The screening parameter, $\kappa$ is related to the Debye screening parameter $\kappa_D$ by following expressions,

$$\kappa = \frac{\kappa_D}{\left[1 - (\kappa_D\sigma)^2/2 + (\kappa_D\sigma)^3/6\right]^{1/2}} \tag{36}$$

$$\kappa_D = \left( \frac{4\pi}{\varepsilon k_B T} \sum_\alpha \rho_\alpha q_\alpha^2 \right)^{1/2} \qquad (37)$$

The above solution of the ionic pair correlation function with the renormalized screening length has been found to be accurate at moderate and high concentrations for binary electrolytes. Thus it considerably extends the range of validity of the classical Debye-Huckel theory [31-32]. We then obtain the Fourier transform of the pair correlation function as,

$$h_{\alpha\beta}(k) = -\frac{4\pi q_\alpha q_\beta}{\varepsilon k_B T(1+\kappa\sigma)} \frac{\kappa^2}{\kappa_D^2(k^2+\kappa^2)} \left( \cos k\sigma + \frac{\kappa}{k} \sin k\sigma \right) \qquad (38)$$

We next combine Eqs. 18, 26-33, and 38 to obtain the following final expression for the microscopic electrolyte friction:

$$\zeta_{elec}(t) = \frac{k_B T q_s^2}{3(2\pi)^3} \int d\mathbf{k}\, k^2 c(k)^2 F_s(k,t) \times \sum_{\alpha,\beta} \sqrt{\rho_\alpha \rho_\beta}\, q_\alpha q_\beta G_{\alpha\beta}(k,t) \qquad (39)$$

In the present study $F_s(k,t)$ is assumed to be equal to $e^{(-D_s K^2 t)}$, as we are interested in the zero-frequency friction. The quantity $c(k)$ used in Eq. 40 is given by,

$$c(k) = -\frac{4\pi}{\varepsilon k_B T(1+\kappa\sigma)} \frac{\kappa^2 \Delta(k)}{\kappa_D^2(k^2+\kappa^2)} \left( \cos k\sigma + \frac{\kappa}{k} \sin k\sigma \right) \qquad (40)$$

## III. Numerical results and discussion

In this work we particularly study the representative of a fully dissociated electrolyte solution, KCl varying its concentration up to 1 M. In the present numerical calculations, all ions in the solutions are assumed to be of equal diameter, $\sigma$ for simplicity. The solution is considered to be of a symmetric salt such that if the charge of each positive ion is $q_1$, then that of each negative ion is $-q_1$. We have also assumed that $\rho_1 = \rho_2$ and $D_1 = D_2$. The solutions can be completely specified by denoting the values of charge and ionic diameter of one chloride ion at temperature, T=298K. The dielectric constant, $\varepsilon$ for all solutions are kept fixed at 80. For high ionic concentrations, Eq. 12 and inverse Laplace transforms of Eq. 39 are to be solved self-consistently by evaluating the wave-vector integrals numerically.

To initiate with the self-consistence scheme in the MCT calculation we exploit the self-diffusion coefficient value of the infinitely diluted KCl solution. This is evaluated from long time molecular dynamics simulation with explicit water molecules. Again, because of the long time tail in the relaxation phenomena, classical molecular dynamics (MD) with explicit water molecules are often found to be computationally expensive specifically at lower k value limits and often fails to characterize the self-diffusion coefficients properly. Thus we used Brownian dynamics (BD) simulation at the Smoluchowski level of approximation to analyze such long time relaxation behaviour. This method indeed is able to calculate exactly the Kirkwood integral. Moreover the computational cost has been improved by incorporating implicit solvent environment in the electrolyte solution. The details of system set-up and simulation method are discussed in the Supplementary Material (see supplementary material [58]). In the present study we also illustrate and compare the essential BD and MD simulation results in connection with MCT.

## A. Time and concentration dependence of self-diffusion coefficient

Using MCT approach we evaluate the time dependence of diffusion coefficient, $D(t)$ shown in **Figure 1**. For small times the diffusion coefficient is close to the value of infinitely diluted solution. The relaxation effect decreases this transport coefficient only at longer time and converges to the characteristic value of self diffusion coefficient of the corresponding ion. We assessed that the MCT result is in excellent agreement with that of BD simulation especially at longer time.

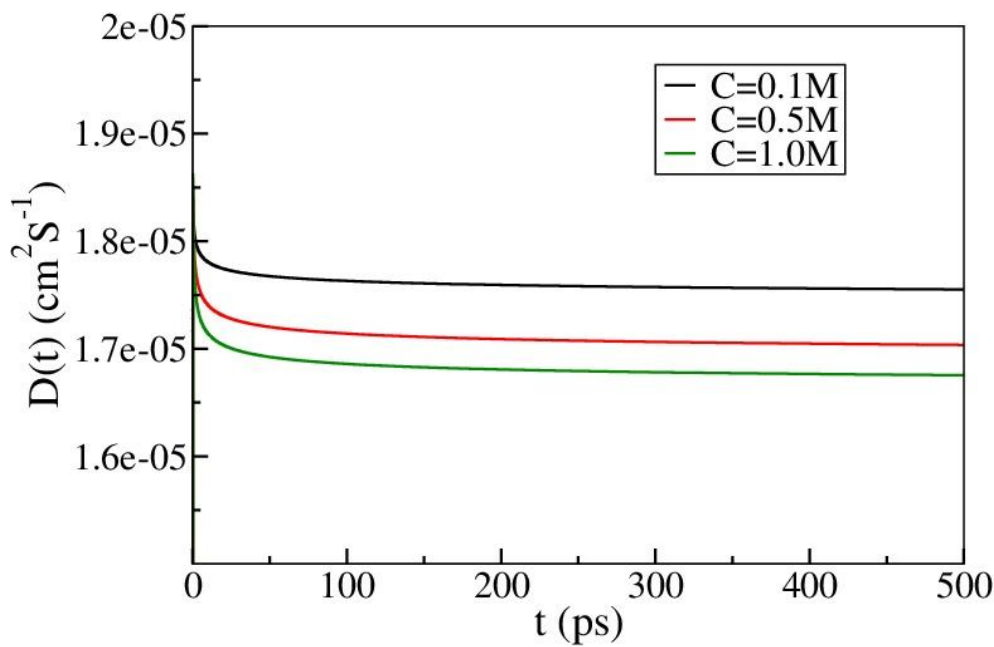

**Figure 1:** Time dependence of self-diffusion coefficient of $Cl^-$ for C=0.1M, C=0.5M and C=1.0 M KCl solutions. For t→0 limit the self-diffusion coefficient is close to the infinitely diluted solution value. Note the convergence of the dynamics at long time to a characteristic value which refers to the self diffusion coefficient of $Cl^-$ ion in this electrolyte solution.

The above excercise helps us to calculate the self-diffusion coefficients at different concentrations of the ionic solution. This particular analysis shows that our microscopic theory provides a satisfactory description of the concentration dependence of self-diffusion coefficient even at higher concentration limit. Thus **Figure 2** illustrates the robustness of our theoretical construction and shows a $\sqrt{C}$ dependence of the self-diffusion coefficient. It is also clear from this figure that our theoretical predictions are in excellent agreement with early experimental results [59]. The correlation with the present Brownian dynamics simulations and molecular dynamic simulation results at low concentration limit found quite satisfactory while at high concentration limit (~1M), the deviation between MCT and simulations is nominal.

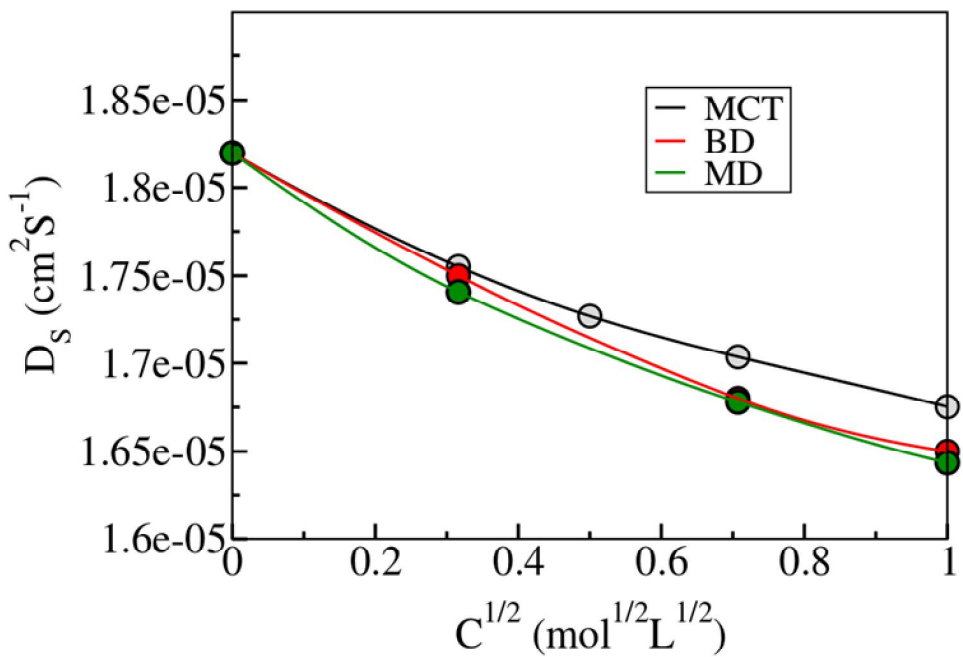

**Figure 2**: Concentration dependence of self-diffusion coefficient of $Cl^-$ ion in KCl solution and comparison among the same evaluated from mode coupling theory (MCT), Brownian dynamics (BD) and molecular dynamics (MD) calculations and analyses. The values of self-diffusion coefficients extracted

from MCT are presented by black circles and they are connected by a black solid line to show their concentration dependent variation as found from self-consistent MCT. The agreement in the calculations of self-diffusion coefficient from MCT, BD and MD is found quite satisfactory especially at lower concentration regime. A slow deviation of BD and MD calculation from MCT is followed as we increase the concentration. However all the connected curves show a weak $\sqrt{C}$ dependence.

**B. Frequency dependence of electrolyte friction**

In an early study Chandra, Wei and Pattey (CWP) derived frequency dependent conductivity under appropriate limiting situation [60]. However for many complex system CWP theory was found to be inadequate which, in part, can be attributed to the absence of full self-consistent calculation of frequency dependence of electrolyte friction. In order to study frquency dependence of electrolyte friction, in **Figure 3** we have plottted ζ(z) vs. z at different ionic concentrations of the electrolyte solution which exhibits a profound concentration dependence. We also observe that with increasing ion concentration, the dispersion of the electrolyte friction is found to occur at a higher frequency because of the faster relaxation of the ion atmosphere. It is worth mentioning here that while inverse relaxation time determine the dispersion of microscopic electrolyte friction, frequency dependence of ionic conductivity is determined by frequency dependent diffusion coefficient D(z). The primary dispersion of D(z) occurs at a much higher frequency than that of the electrolyte friction because of the presence of a frequency term in generalized Einstein relation that eventually connects the friction and diffusion at finite frequencies.

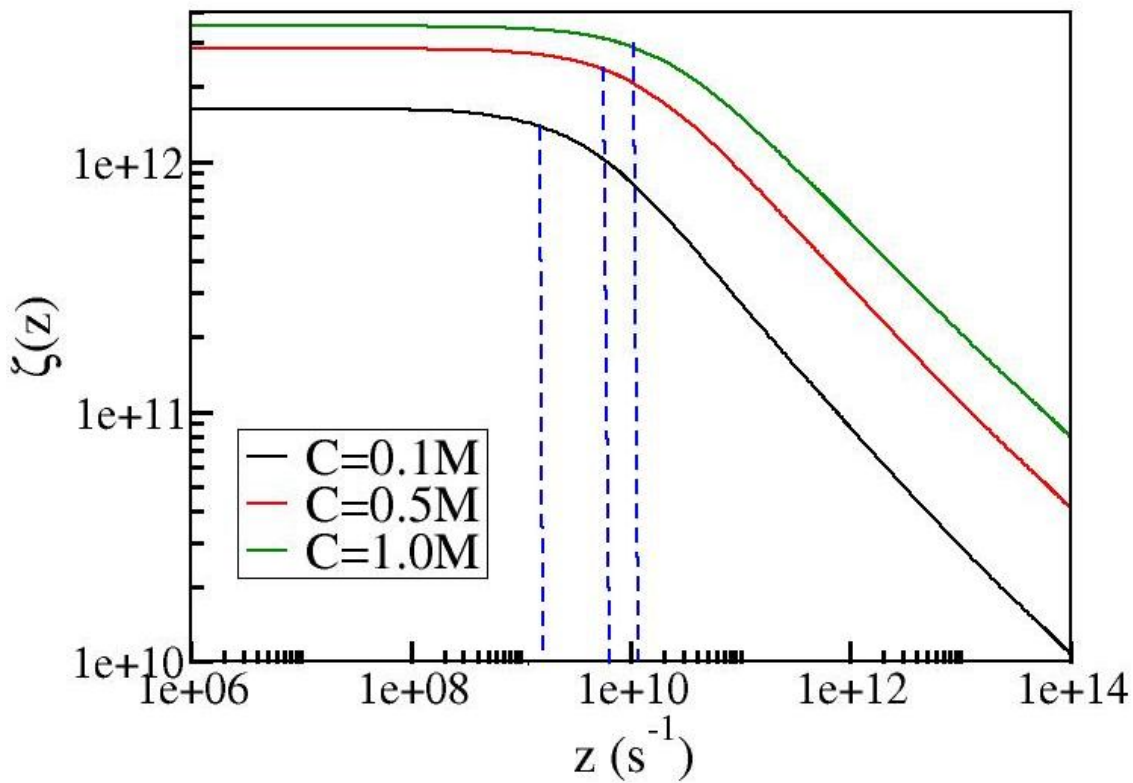

**Figure 3**: The frequency dependence of electrolyte friction of an ion at different ionic concentrations of eletrolyte solution: (a) C=0.1M, (b) C=0.5M, and (c) C=1.0M. Note that the dispersion of the electrolyte friction begins at much high frequency with increasing ion concentration. The dispersion for 0.1M and 0.5M solution is found to occur in the nanosecond timescale range.

## C. Wave number dependence of dynamic structure factor with varying concentration

Since different wavenumber (k) values correspond to different length scales, we can have information about the way in which the relaxation process of a system differs with different observing length scales from a k-dependent F (k, t) study at different concentrations of the ionic solution[61]. **Figure 4(a)** illustrates that for low concentrations of KCl solution (C=0.1M), F(k,t) exhibits a clear exponential relaxation behaviour at relatively larger wave number (k=0.1 Å$^{-1}$). As we lower the k values the relaxation behaviour turns out to be bi-exponential for the same concentration limit (see **Figure 4(b)**). The agreement between BD and MD calculations of dynamic structure factor is again found to be excellent for lower concentrations of this electrolyte solution. However, for higher concentrations (towards 1M), in the faster component of the relaxation time while we do not find very good agreement, at longer timescale region F(k,t) relaxation from MCT and BD shows excellent convergence (see **Figure 4(c)**).

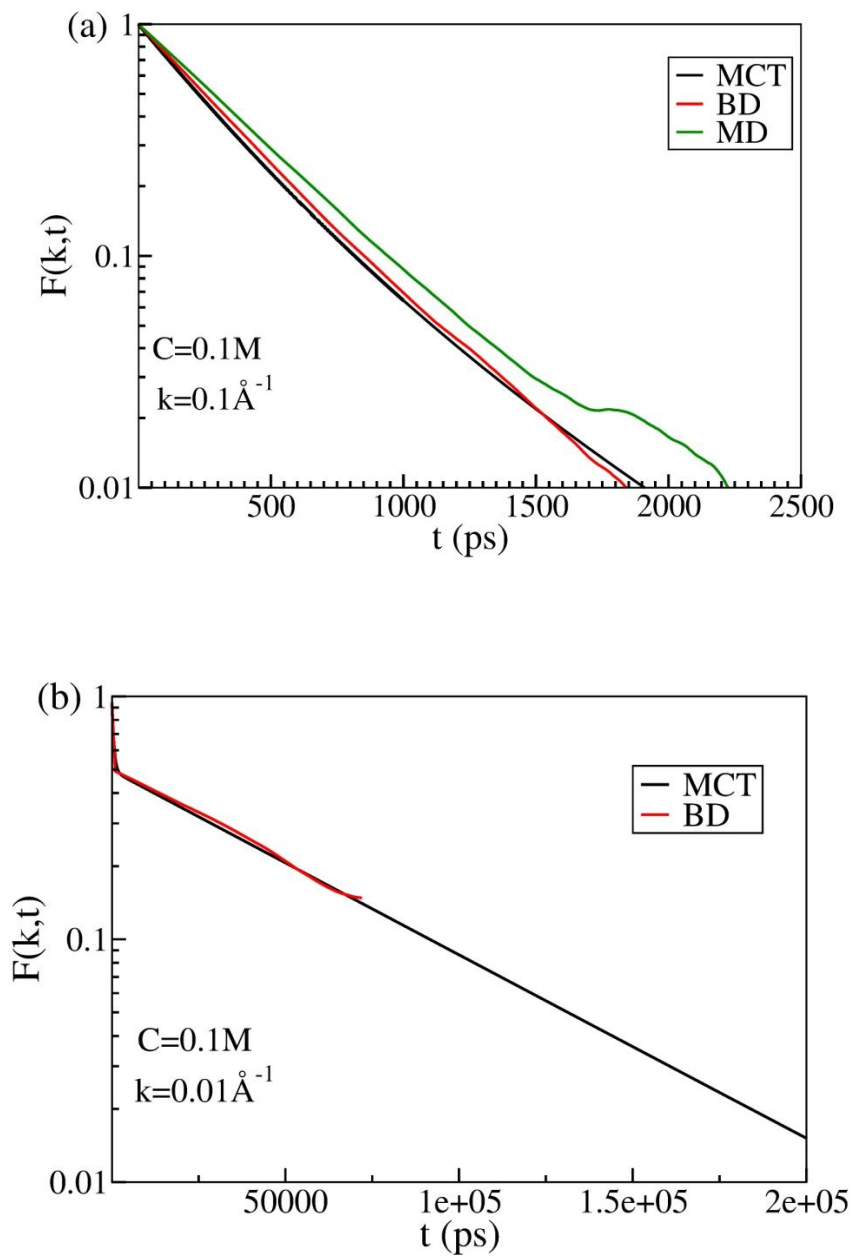

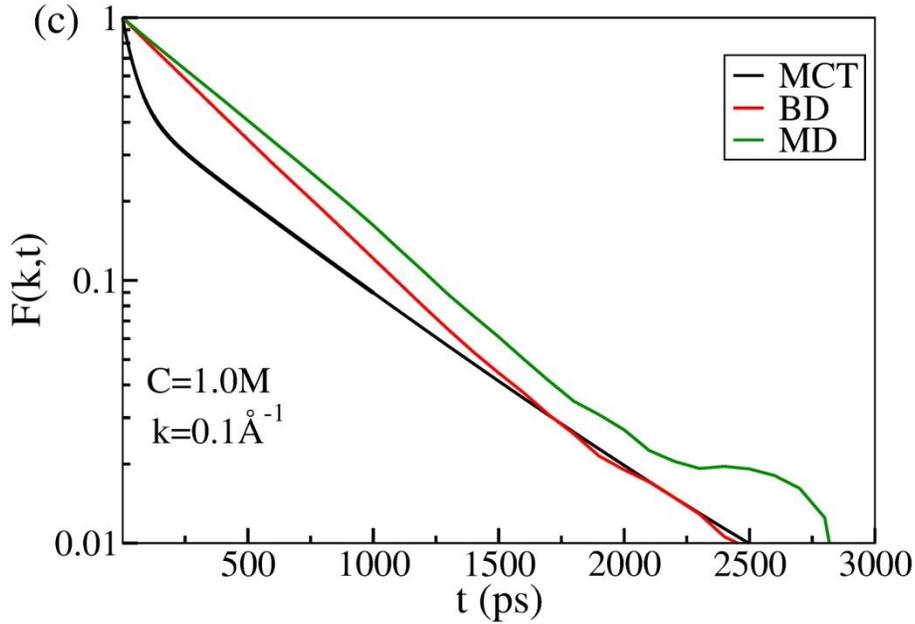

**Figure 4:** Ion-ion dynamic structure factor (F(k,t)) extracted from mode coupling theory (MCT), Brownian dynamics (BD) and molecular dynamics (MD) calculations and analyses. (a) F(k,t) at 0.1M KCl, and at wave number k=0.1Å$^{-1}$. Characteristic exponential long time decay with excellent aggrement between BD and MCT are evident at this low concentration regime. (b) F(k,t) at 0.1M KCl, and at lower wave number, k=0.01Å$^{-1}$. Long time biexponential decay kinetics again with a fine convergence between MCT and BD calculations. (c) F(k,t) at higher concentration limit i.e., at 1M KCl, and at wave number, k=0.1Å$^{-1}$. The aggrement between MCT and BD is found to be quite satisfactory especially at the longer time region. All the relaxations evaluated from MD calculation follow a rather slower decay kinetic than the other two.

When we plot the temporal decay of F(k, t) in a log-log scale at several k values, we observe a significant non-exponential relaxation behaviour with power law decay regimes separated by an intervening plateau at intermediate timescale range with decreasing k. This imparts a step-like feature to the temporal behaviour of F(k,t). As we increase the concentration the low k effect becomes more prominent (see **Figures 5(a), (b), (c)**). This can be interpriated as follows: At large k, the length scale become very small in the range of ionic diameter. Thus the collective fluctuation does not play a significant role in the relaxation of F(k,t), leading to an exponential relaxation behavior. On contarary, at small k values, the length scale being significantly large, the collective fluctuations play dominant role in the relaxation of F(k,t) which is manifested as a power law decay. It seems that a cooperative phenomena emerges in the collective dynamics of the system. Such power law relaxation near small wavevector is an area of great interest in the recent years.

The wave number dependence of F(k,t) relaxation reveals a striking similarity to the temperature dependent relxation of the similar dynamic quantity for supercooled liquids as the glass transition is approached [62-64]. For the supercooled liquid the emergence of such step-like feature is well understood as a consequence of β-relaxation. This analogous behaviour signifies a characteriztic dynamical feature appearing as a dramatic retardation in the movement of super-cooled liquid at lower temperatures and ionic liquid at lower wave numbers. For super-cooled liquid this has been attributed to the cage-effect, the lack of vacancies in the first-neighbour shell surrounding a particle inhibiting large displacements. In the present study, this might be originating as a consequence of long range electrostatic interactions between tagged ion and the surrounding environment.

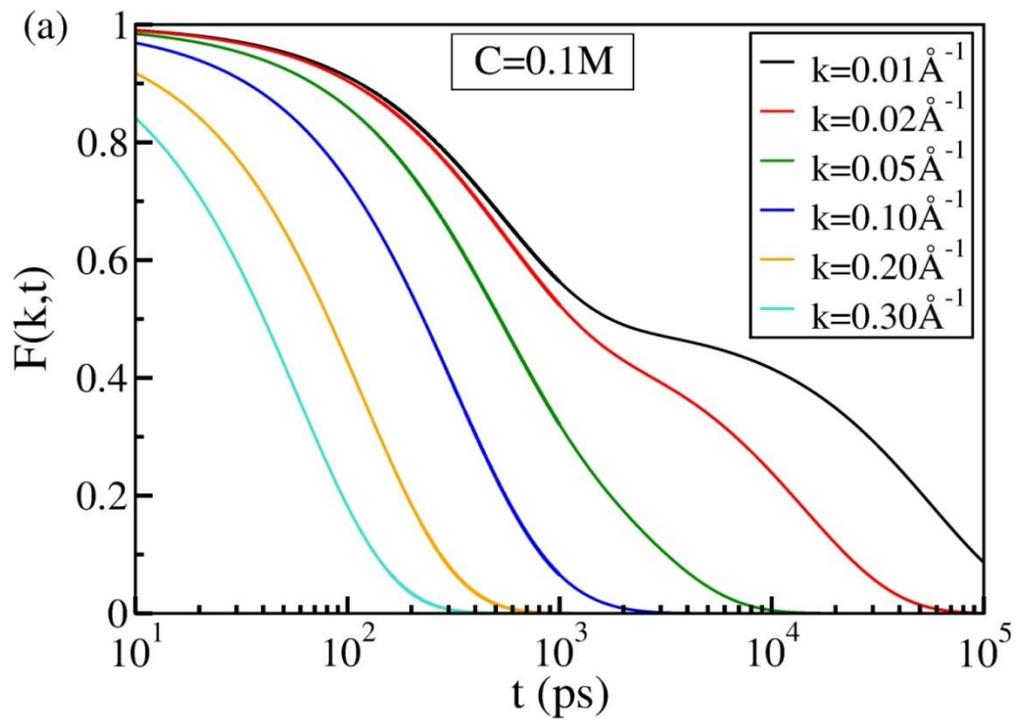
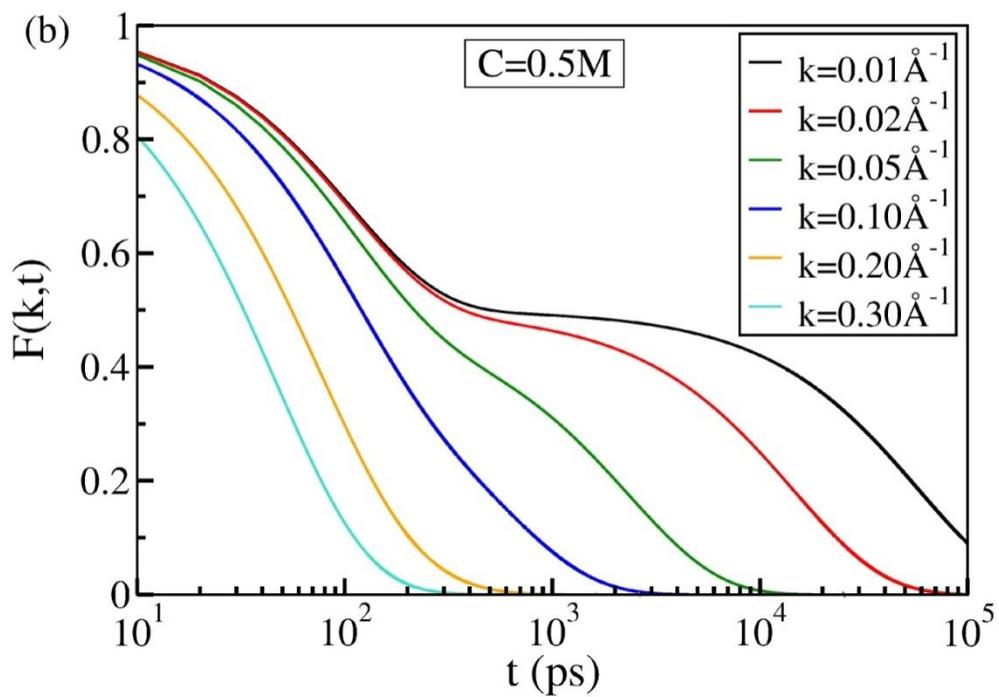

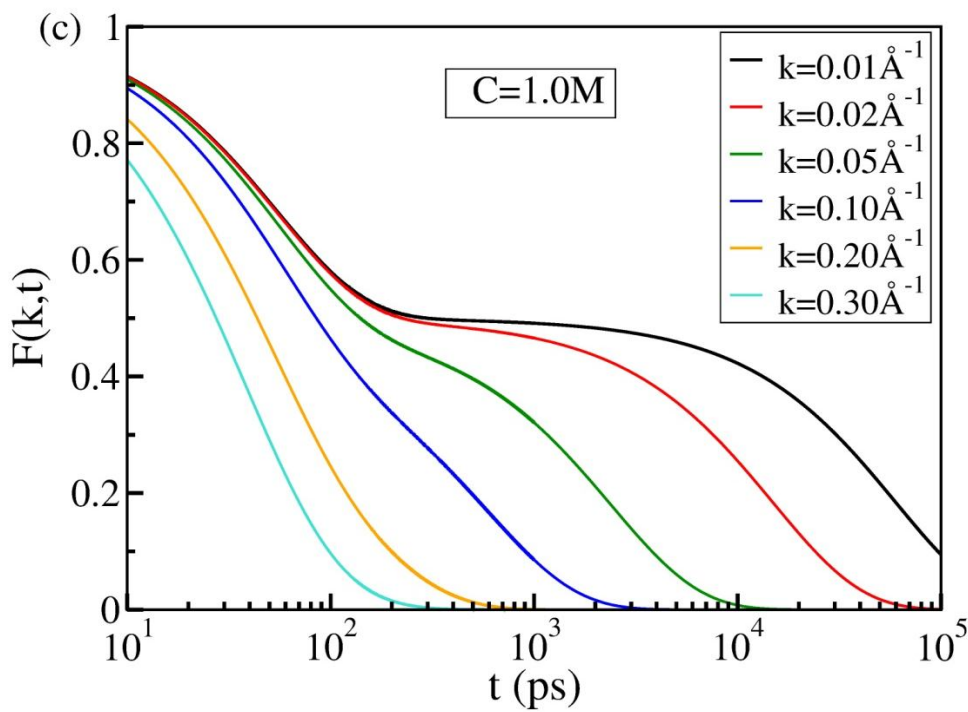

**Figure 5**: Wave number dependence of ion-ion dynamic structure factor (F(k,t)) for Cl$^-$ ion in (a) 0.1M (b) 0.5M (c) 1.0M of KCl solution, in semilog scale. Note the distinct ladder-like relxation at small wave number limit which exhibits a striking similarity with supercooled liquids.

It is important to mention here that while at small k values (below k=0.1 Å$^{-1}$), the decay of dynamic structure factor can be described by a power law at intermediate timescale (beyond 100ps), along with a slow non-exponential relaxation dynamics (as shown in **Figure 5**), this clear power law behaviour eventually disappears at large k values, beyond k=0.05 Å$^{-1}$.

### D. Analysis of solvation time correlation function, $C_S(t)$

The main hypothesis of the present work is that a highly non-exponential solvation dynamics can emerge in the long time because of the long range nature of ion-ion interaction. This long range nature, in turn, makes the contribution of the small wavenumber part of dynamic structure factor (or, intermediate scattering function) to solvation dynamics significant. That is, a tagged ion or a dipole with a distributed charge can experience fluctuations of ion atmosphere that is very large. Relaxation of such a large ion atmosphere can be slow. According to the mode coupling theory, the solvation time correlation function derives contribution from a range of wavenumbers, that is, from ion atmospheres of different lengths or sizes. Since the relaxation rates of these ion atmospheres depend strongly on size through the k-dependence of the intermediate scattering function, the solvation TCF in turn becomes highly non-exponential. We have seen that F(k,t) develops a well-defined plateau due to the separation of time scales between the initial and the long time decay. This separation becomes more pronounced at smaller wave numbers. This separation of time scales between the initial and final parts of decay further accentuates the non-exponentiality of the solvation TCF and can even give rise to a power law decay, as discussed below with numerical results.

**(i) Wave number based analysis from MCT**

As discussed above, **Figure 5** demonstrates the appearance of a plateau appears in the relaxation of the dynamic structure factor at low wave number limit, particularly in the range of k<0.05Å.

We now numerically test the validity of the proposition that the nonexponential or power law solvation dynamics arises from the long wavelength ion-atmospheric relaxation. In **Figure 6,** we show the relative contributions from different range of k by separating the integration of Eq. 6 in the following fashion

$$<\delta E(0)\delta E(t)> = A\int_0^\infty dk\, k^2\, c_{pi}^2(k)\, F_{ij}(k,t)$$

$$= A\left[\int_0^{k_s} dk\, k^2\, c_{pi}^2(k)\, F_{ij}(k,t) + \int_{k_s}^\infty dk\, k^2\, c_{pi}^2(k)\, F_{ij}(k,t)\right] \quad (41)$$

The 1st integration in Eq. 41 contains the contribution of the small wavenumber range and the 2nd term contains the contribution of the intermediate to large wavenumber range. While separation in Eq.41 is exact, we shall use $k_s$ as a parameter or marker that separates small from intermediate in the wave number space.

Thus in **Figure 6,** we show the relative contribution from different k domains with $k_s$= 0.05Å$^{-1}$.

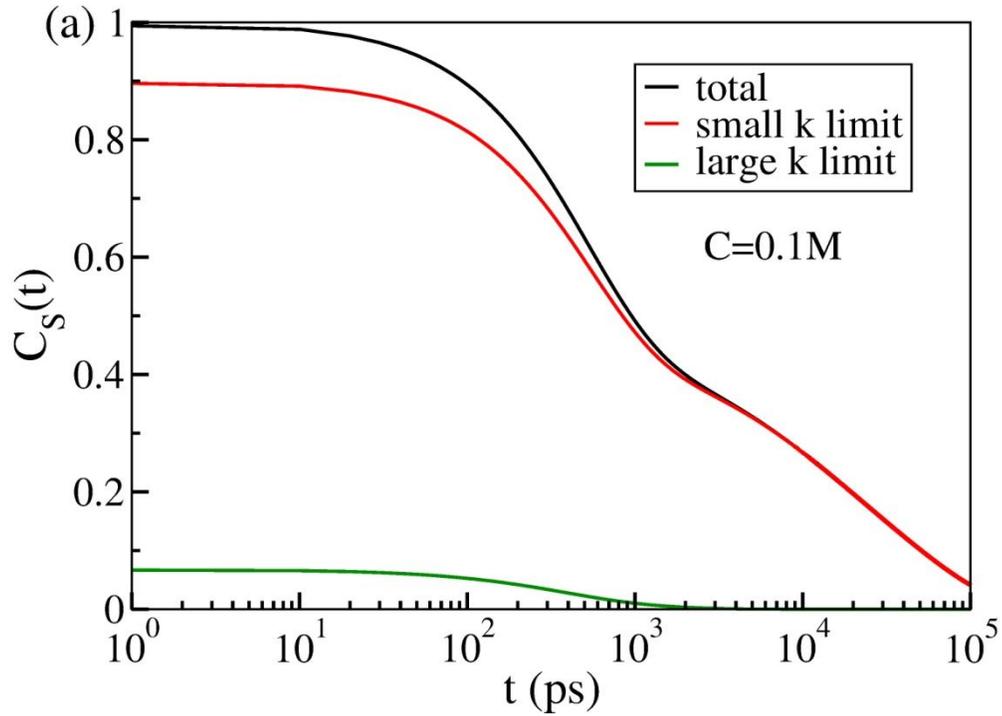

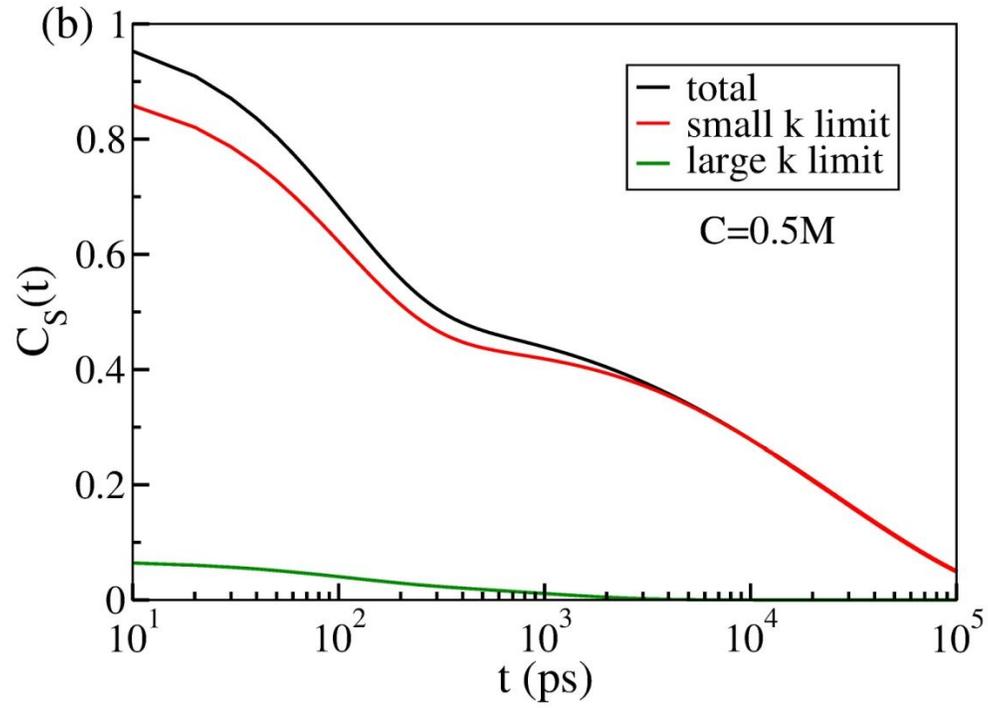

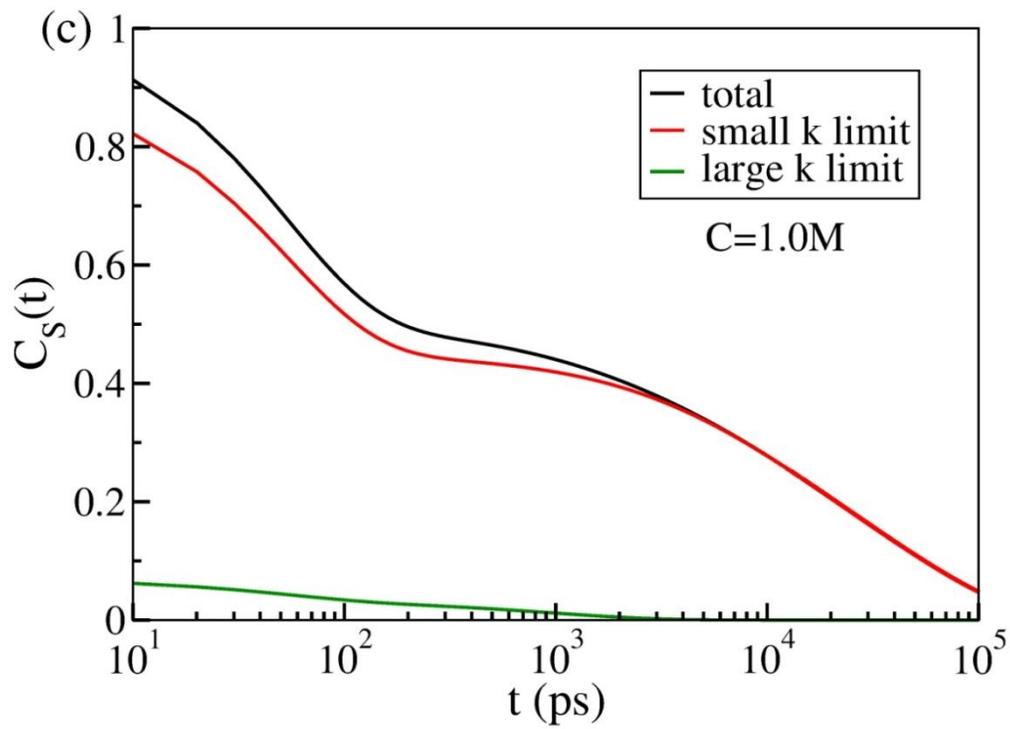

**Figure 6:** Time dependence of wavenumber (k) contribution to the total solvation time correlation function (STCF) derived from MCT approach at three different ionic concentrations: (a) C=0.1M, (b) 0.5M and (C) C=1.0M. For all these concentrations, $C_S(t)$ for large wave number variation shows an exponential decay and imparts a smaller contribution (< 10%) only to the faster relaxation component of the total STCF. Response from the low wavenumber variation is significantly large (> 85%) to the total STCF. Total STCF shows a stretched exponential decay with significantly large nonexponential character.

We find that the time dependence of this contribution to the total solvation time correlation function is largely coming from the small wave number (< 0.05Å$^{-1}$) variation. While for the large k, the STCF decays exponentially, STCF for low k variation are best fitted by a stretched exponential functions of time with stretching exponents $(\beta)$ in the range of 0.58-0.62. The expression used for the best fit is as follows:

$$C_S(t) = C_1 e^{-(t/\tau_1)} + C_2 e^{-(t/\tau_2)^\beta} \tag{42}$$

**Table 1** contains a summary of the parameters from stretched-exponential fits. In general, the values of β observed here endow with a significant nonexponential nature to the STCF.

**Table 1:** Parameters of the stretched exponential fit to the average solvation time correlation function (evaluated from **Figure 6**).

| $C_{KCl}$ | $C_1$ | $\tau_1$ (ns) | $C_2$ | $\tau_2$ (ns) | β |
|---|---|---|---|---|---|
| 0.1M | 0.53 | 0.52 | 0.47 | 24.7 | 0.62 |
| 0.5M | 0.49 | 0.11 | 0.51 | 24.2 | 0.59 |
| 1.0M | 0.48 | 0.05 | 0.52 | 23.4 | 0.58 |

**(ii) Solvation time correlation function from BD simulation**

As we mentioned in the introduction that in the DNA solvation dynamics experiments by Berg and coworkers, the probe was an optically excited coumarin [41-44]. Moreover the DNA was immersed in a buffer solution. On the other hand, in the experiments of Zewail et al. the probe was 2-aminopurine dye [46]. In the absence of an elite technique of obtaining direct correlation function, we have used albeit a simple procedure of tagging ions one by another and conducted Brownian dynamics (BD) simulation for each tagged ion.

The average solvation time correlation function extracted from our BD simulations results weak power law behaviour because of self motion of the ions. This self motion is known to be effective at longer time. In **Figure 7** we show the clear emergence of the power law.

In the solvation dynamics experiments of aqueous electrolyte solution it is observed that the relaxation of ion-solution interaction energy can be separated into its ionic components and the associated solvent component. Eventually, the ionic relaxation has been found to be significantly slower than the solvent response.

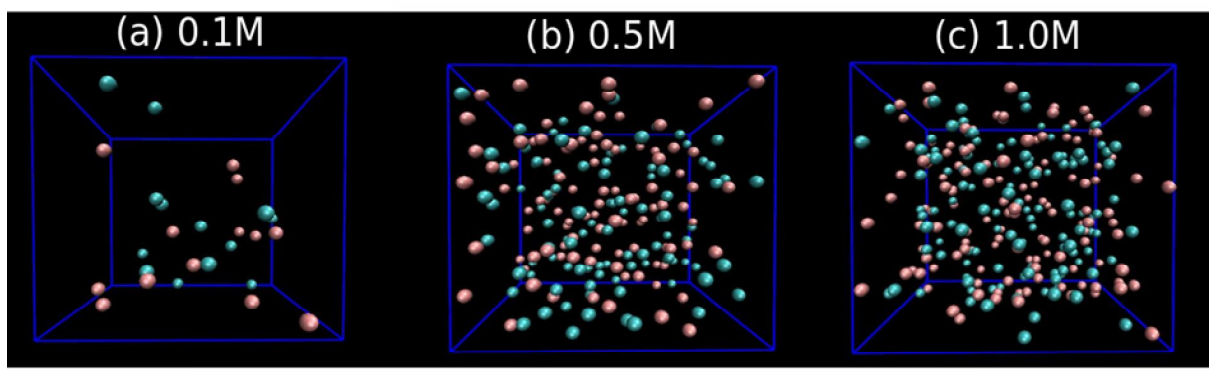

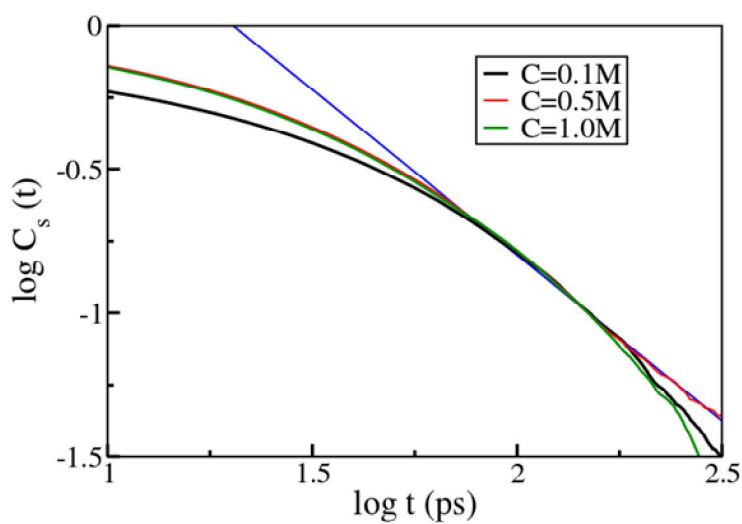

**Figure 7**: Solvation response of tagged Cl$^-$ ions in aqueous KCl solution along with the snapshots extracted from their equilibrium trajectory at different ionic concentrations: (a) 0.1M, (b) 0.5M and (c) 1.0M, where cyan particles are Cl$^-$ ion and pink particles are K$^+$ ion. Time dependence of the total solvation energy time correlation functions (TCF) are shown below for these different KCl concentrations. The long time decay of the solvation TCF decreases with ion concentration. In all these cases an emergence of power-law behaviour is evident (marked by blue straight line) that sets beyond ~35ps range with power law exponent ~1.02.

**E. Concentration dependence of dynamic structure factor**

In order to deepen our understanding of the role of electrolyte dynamics in the solvation process, we have also investigated the concentration effects on the dynamic structure factor (see **Figure 8**). Conventionally, the more the concentration, the slow is the long time decay. However, intrestingly we find that the relaxation behaviour is somewhat different at the fast to intermediate time scale region. We believe this provides the microscopic explanation of the weak concentration dependence of the dynamical properties of the ions at that time limit.

Convergence in the long time decay to a common relaxation curve shown in **Figure 8** is indeed surprising. This seems to happen rather sharply around 1ns time scale region where STCF has attained a value of ~0.4. The deacy is such that $C_S(t)$ varies as $a - b\log(t - t_0)$, where $t_0$ is approximately 1ns. Note that Berg and coworkers [41-44] also observed similar closeness in the logarithmic decay.

The remarkable insensitivity of the STCF to the concentration at longer time has also been observed in the BD simulation in around similar time range (see **Figure 7**). Such an insensitivity to the concentration reflects the weak dependence of the electrolyte friction to the wavenumbers, as shown in the inset of **Figure 8**. Small wavenumbers, however, makes the dominant contribution to the STCF as observed earlier in **Figure 6**. Such weak concentration dependence seems to imply there is a cancellation between the concentration dependence of solvation frequency and the solvation friction terms [65, 66].

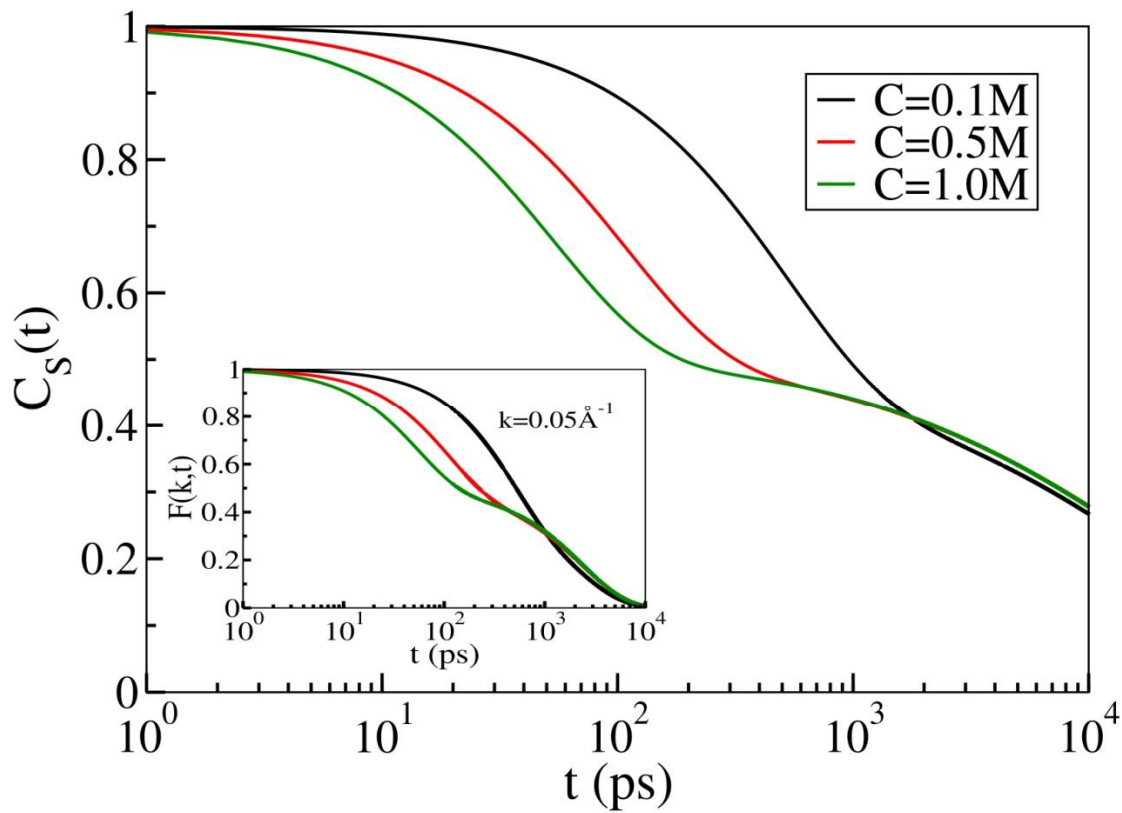

**Figure 8**: Concentration dependence of solvation time correlation function (STCF) along with the dynamic structure factor shown in the inset where k=0.05Å$^{-1}$ for each case. The cross over between higher and lower concentrated solutions is observed at intermediate time scale range. Note the insensitivity of the STCF to the concentration at longer time.

The above result, particularly the concentration dependent crossover could be understood in terms of a complex interplay between structure and non-Markovian dynamics, that is, frequency dependent friction. As concentration decreases, collective ion dynamics should become slow. However, k-dependence complicates the matter because structure moves to lower k as concentration is lowered.

## IV.     Summary and Conclusions

A self-consistent theory of time dependent diffusion, dynamic structure factor and solvation time correlation function of an electrolyte solution is presented in this work. The theory is based on ideas of mode coupling theory. This approach provides valuable insight into electrolyte dynamics in leading towards a microscopic picture that can be used to understand and to extend the well-known phenomenological laws of electrochemistry.

The theory combines the mode coupling approach and time dependent density functional approach. The additional MCT slow variables here are the charge density and the current density. The details of the static and dynamic ion-ion correlations and also the effects of self-motion of the ions have also been incorporated in this construction. The prime effects of ion atmosphere relaxation are included which is now frequency dependent. This treatment is developed with the frequency dependent electrolyte friction and the diffusion those are found to be dynamically coupled and solved self-consistently.

The present study explains the reason for the apparently different observations made by two different groups on DNA solvation. By studying carefully we find that there are actually no significant differences between the two because Zewail et al. used a much shorter time range as they employed a Laser Pulse of 150fs resolution [46]. Therefore, the experiments of Zewail et al. explored much shorter timescales, in the range of picosecond only that bears a close resemblance to the decay constant of our STCF calculations, at larger concentrations (C=0.5M ). Note that our times are overestimated as we have employed a BD approach. This, in turn, implies that we have missed the shorter time decay which could very well occur with the time constants in the range of few tens of picosecond or even less. Berg et al., on the other hand, used a much larger time window which led to the discovery of the power law relaxation[41-44]. The present theoretical investigation entirely explains all the above experimental findings. Earlier MD simulation studies captured largely the initial decay.

The present study comes up with several other interesting findings. First, the time dependence of ionic diffusion coefficient in combination with molecular dynamic simulation results can popularize the use of such molecular theory to understand limiting situations of the dynamics. While MCT, BD and MD simulations, all the treatments help us to achieve a converged value of self-diffusion coefficient for lower ionic concentrations from longer time evaluation, the merger seems to deviate at larger concentrations. In the short time limit, the self-diffusion coefficient is close to the infinitely dilute solution value. The relaxation effect decreases this transport coefficient only for times greater than the Debye relaxation time. The inconsistency in the measurement of self-diffusion coefficient from time of flight neutron scattering and NMR or tracer methods can be explained by this observation [67]. While ion-atmosphere would take longer time to relax, the time scale of such neutron scattering experiments is 20 ps, which is indeed much less than ion-atmospheric relaxation time. Consequently neutron scattering experiments do not account the whole relaxation effect. The long time methods such as NMR would always be recommended for such measurements.

From this study we may surmise that at long times the hydrodynamic interactions are not important for the diffusion of small sized ions. The concentration dependence of self-diffusion coefficient evaluated from this theory unifies other phenomenological approaches. With increasing concentration, the dispersion of electrolyte friction is found to occur at a higher frequency due to the faster relaxation of the ion atmosphere. The wave number dependence of dynamic structure factor $F(k, t)$, describes dynamical relaxation at different length scales. Such behaviour has drawn an intriguing analogy with temperature dependent relaxation dynamics of supercooled liquids. The emergence of the power-law-form has been successfully tested here from the solvation dynamics study probing each ion. This ion-ion dynamic structure factor indeed exhibit temporal power law decay at intermediate times. Our observation of the concentration dependence of *F(k,t)* is that the more concentrated electrolytes relax at longer time while showing faster decay at short time scale limit. The dynamical crossover that we demonstrate between less concentrated and more concentrated electrolytes has also been observed in the evaluation of velocity time correlation function from early Brownian dynamics and MCT calculations[59]. However, the appropriate explanation for such behaviour remains unclear.

Although the present theory incorporates static and dynamic ion-ion correlations and the screening effects, the molecular details of the ion-solvent and solvent-solvent correlations are still missing. One can now investigate how the presence of a shell of water molecules solvating the ion, and its dynamics accompanying the motion of the ion, contributes to the electrolyte friction and conductance[33, 68-71]. Of course, such a study involves more complexities and requires a more efficient microscopic approach in which the solvent molecules are treated explicitly. Inclusion of molecularity of the solvent in the calculation of electrolyte friction is a rather daunting task. We hope to address some of these issues in future.

## Acknowledgment

We thank Dr. Parveen Kumar for many useful discussions. This work was supported in parts by grants from DST, India. BB acknowledges support from JC Bose fellowship from DST, India.